%% file: nstarsII_final.tex
\documentclass[10pt,preprint2]{aastex}

\shorttitle{Nearby Stars}
\shortauthors{Gray et al.}

\begin{document}

\title{Contributions to the Nearby Stars (NStars) Project:
  Spectroscopy of Stars Earlier than M0 within 40 parsecs: The
  Southern Sample}
\author{R.O. Gray}
\affil{Department of Physics and Astronomy, Appalachian State
  University, Boone, NC 28608}
\email{grayro@appstate.edu}
\author{C.J. Corbally}
\affil{Vatican Observatory Research Group, Steward Observatory, 
Tucson, AZ 85721-0065}
\email{corbally@as.arizona.edu}
\author{R.F. Garrison}
\affil{David Dunlap Observatory, P.O. Box 360, Station A, Richmond
  Hill, ON L4C 4Y6, Canada}
\email{garrison@astro.utoronto.ca}
\author{M.T. McFadden, E.J. Bubar\altaffilmark{1} and C.E. McGahee }
\affil{Department of Physics and Astronomy, Appalachian State
  University, Boone, NC 28608}
\altaffiltext{1}{Department of Physics and Astronomy, Clemson
  University}
\author{A.A. O'Donoghue and E.R. Knox}
\affil{Dept of Physics, St. Lawrence University, Canton, NY 13617}

\begin{abstract}
We are obtaining spectra, spectral types and basic
physical parameters for the nearly 3600 dwarf and giant stars earlier
than M0 in the {\it Hipparcos} catalog within 40pc of the Sun.  Here we 
report on results
for 1676 stars in the southern hemisphere observed at Cerro Tololo
Interamerican Observatory and Steward Observatory.  These results
include new, precise, homogeneous spectral types, basic physical
parameters (including the effective temperature, surface gravity, and
metallicity, [M/H]) and measures of the chromospheric activity of
our program stars.  We include notes on astrophysically
interesting stars in this sample, the metallicity distribution of the
solar neighborhood and a table of solar analogues.  We also
demonstrate that the
bimodal nature of the distribution of the chromospheric activity parameter 
$\log R^{\prime}_{\rm HK}$ depends strongly on the metallicity, and we
explore the nature of the ``low-metallicity'' chromospherically active
K-type dwarfs. 
\end{abstract}

\keywords{Galaxy: solar neighborhood---stars: abundances---stars:
activity---stars: fundamental parameters---stars: late-type---stars: 
statistics}

\section{Introduction}
This is the second in a series of three papers which present results
of a joint study of the nearby solar-type stars under the aegis of the   
NASA/JPL Nearby Stars/{\it Space Interferometry Mission} Preparatory
Science Program.  The goal of this project is to obtain spectroscopic
observations of all 3600 main sequence and giant stars with spectral
types earlier than M0 in the {\it Hipparcos} catalog \citep{esa97} within 
40pc of the Sun.  We have obtained
blue-violet classification-resolution (1.5 - 3.6\AA) spectra for all
of these stars to date.  These spectra are being used to obtain
homogeneous, precise, MK spectral types.  In addition, these spectra
are being used in conjunction with synthetic spectra and existing
intermediate-band Str\"omgren $uvby$ and broadband $VRI$ photometry to
derive the basic astrophysical parameters (the effective temperature,
gravity, and overall metal abundance [M/H]) for many of these stars.
We are also using these spectra, which include the \ion{Ca}{2} K and H
lines, to obtain measures of the chromospheric activity of the program
stars on the Mount Wilson system.  The purpose of this project is to
provide data that will permit an efficient choice of targets for both
the {\it Space Interferometry Mission (SIM)} and the {\it Terrestrial
  Planet Finder (TPF)}.  In addition, combination of these new data
with existing kinematic data should enable the identification and
characterization of stellar subpopulations within the solar
neighborhood.

We report in this paper on southern hemisphere stars for which
observations have been carried out on the 1.5~m telescope at Cerro
Tololo Interamerican Observatory (CTIO) and the 2.3~m Bok Telescope at
Steward Observatory (SO) which is situated on Kitt Peak.  Other
observations for this project
have been carried out on the 0.8~m telescope of the Dark Sky
Observatory (DSO) and the 1.8~m telescope at the David Dunlap Observatory.
Observations of 664 stars carried out at DSO were the subject of the
first paper of this series \citep{gray03} (hereinafter paper I).  The 
remaining stars in our sample
will be reported on in the third paper of this series (Gray
et al. 2007, in preparation). 

\section{Observations and Calibration}

The observations reported in this paper were made with the 1.5~m
telescope at CTIO and the 2.3~m Bok telescope at Steward Observatory.
The CTIO observations were carried out using the Loral $1200 \times
800$ CCD on the Cassegrain spectrograph.  Grating \#58 was used in the
second order with the CuSO$_4$ order-blocking filter and  a slit size 
of 86$\mu$m to give a nominal resolution of
2.6\AA\ (2 pixels) with a wavelength range of $3800-5150$\AA.
However, in practice, the actual resolution of these spectra is closer
to 3.5\AA.  These spectra were reduced using standard methods using
IRAF\footnote {IRAF is distributed by the National Optical Astronomy
  Observatory which is operated by the Association of Universities for
  Research in Astronomy, Inc. under cooperative agreement with the
  National Science Foundation}.  The Loral CCD on the CTIO 1.5m Cassegrain
spectrograph has a number of blemishes which affected these spectra.
In particular, one blemish affects the $4058-4069$\AA\ region of these
spectra; we carefully positioned the spectra so that this blemish did
not affect the important \ion{Sr}{2} $\lambda 4077$ line used in
luminosity classification.  Another
blemish affected about 50\% of the CTIO spectra, causing a distortion
in the violet half of the G-band.  In total, we had 4 observing runs at
CTIO, each lasting 5 - 7 nights.  Table 1 presents a log of these
observing runs.

\begin{deluxetable}{ll}
\tablewidth{0pt}
\tablecaption{Observing Log}
\tablehead{\colhead{Telescope} & \colhead{Dates}}
\startdata
SO Bok 2.3~m & Dec 21 --- 23, 2000 \\
CTIO 1.5~m   & Feb 4 --- 9, 2001 \\
SO Bok 2.3~m & Mar 11 --- 12, 2001 \\
SO Bok 2.3~m & Apr 8 --- 9, 2001 \\
SO Bok 2.3~m & Jun 1 --- 5, 2001 \\
CTIO 1.5~m   & Aug 2 --- 9, 2001 \\
SO Bok 2.3~m & Sep 8 --- 10, 2001 \\
SO Bok 2.3~m & Nov 24 --- 27, 2001 \\
SO Bok 2.3~m & Jan 1 --- 2, 2001 \\
SO Bok 2.3~m & Jun 16 --- 19, 2002 \\
CTIO 1.5~m   & Jun 23 --- 27, 2002 \\
CTIO 1.5~m   & Dec 10 --- 15, 2002 \\
SO Bok 2.3~m & Mar 11 --- 13, 2003 \\
\enddata
\end{deluxetable} 

The Steward Observatory observations were carried out using the
Boller \& Chivens spectrograph with the ccd20 CCD ($1200 \times 800$ pixels) 
on the Bok 2.3~m telescope.  The spectrograph
was used with the 600~g/mm red-blazed grating in the second order, the 
Schott 8612 order blocking filter and a slit 
size of 2.5$^{\prime\prime}$ to give a resolution of
2.6\AA\ and a wavelength range of $3800-4960$\AA.  These spectra were
likewise reduced using standard methods with IRAF.  We had 9
observing runs on the 2.3~m Bok telescope (see table 1).

For stars with spectral types of G8 and earlier, the CTIO and SO
spectra were rectified using an X Window System
program xmk19, written by one of us (R.O.G.) and were used in that
format for both spectral classification and the determination of the
basic physical parameters.  

However, for the late-type stars (G8 and later) rectification is
problematic as no useful ``continuum'' points can be identified.  For
that reason, we have attempted to flux calibrate these spectra even
though they were obtained with a narrow slit.  For this purpose, we regularly
obtained spectra of a number of spectrophotometric standard stars
\citep{hamuy92} during the observing runs at both CTIO and SO.  These
standard observations have been used to remove approximately the
effects of atmospheric extinction and to calibrate the spectrograph
throughput as a function of wavelength.  Except for observations made
at high airmasses, this procedure yields calibrations of {\it
  relative} fluxes with accuracies on the order of $\pm 10$\%.  We
require greater accuracies, however, for the determination of the
basic physical parameters, and so we have applied the technique of
``photometrically correcting'' these fluxes using Str\"omgren
photometry \citep[taken from the {\it General Catalog of Photometric 
Data} by ][]{mermilliod97}.  This technique was described in detail in paper I.
However, because these spectra span only two Str\"omgren bands (as
opposed to the three bands employed in paper I), an additional
flux correction procedure was employed during the analysis to give the
basic physical parameters (see below).

All of the spectra obtained for this paper are available on the
project's Web site.\footnote{http://stellar.phys.appstate.edu}  The
rectified spectra found on that Web site have an extension of .rec.  
Spectra which have been
flux calibrated, but not photometrically corrected, are normalized to
unity at a common point (4503\AA) and have an extension of .nor,
whereas photometically corrected spectra are available in a normalized
format (.nfx) and in terms of absolute fluxes (.flx) in units of erg
s$^{-1}$ cm$^{-2}$ \AA$^{-1}$.  Dates and times of observation and other
information can be found in the ``footers'' of these spectra.

\section{Spectral Classification}

The stars in this paper were classified on the MK system by direct
visual comparison on the computer screen with MK
standard stars selected from the list of ``Anchor Points of the MK
System'' \citep{garrison94}, the Perkins catalog \citep{keenan89} and,
for the late K- and early M-type dwarfs, \citet{henry02}.  Spectra of an
extensive set of relevant standards were obtained on each telescope.  A list of
the standards used (and the actual spectra) can be found on the
project Web site.

Paper I described the techniques we used to carry out the spectral
classifications, and how we are ensuring the homogeneity of our
spectral types, which we derive from spectra obtained
on four different spectrographs.  In short, this homogeneity is
maintained by significant overlaps between the samples observed with
the four telescopes employed by this project, as well as having a
number of MK standards in common between the different samples.  Out
of the 1676 stars reported in this paper, 78 stars were
observed at both CTIO and SO.  Spectral types were assigned
independently for the two groups of stars.  For the 78 stars in
common, we find the following systematic difference in the spectral types: 
\begin{displaymath}
{\rm ST_{CTIO} - ST_{SO}} = 0.19 \pm 0.50
\end{displaymath}
and thus the CTIO spectral types are systematically later than the SO
spectral types by 0.2 spectral subclasses, with a scatter (one point) 
of one-half spectral subclass (note, a subclass is the difference,
for example, between a K0 and a K1 star).  For this comparison, we used the 
numbering system of \citet{keenan84} to convert spectral types 
into numerical codes.  We note that this very small systematic
difference in the spectral types indicates a high degree of
consistency between the northern and southern sets of MK Standards
chosen for this work.

Our spectral types are
multidimensional, as they include not only temperature and luminosity
types, but also indices indicating abundance peculiarities and the
degree of chromospheric activity.

As described in paper I, chromospheric activity is evident in our
spectra through emission reversals in the cores of the \ion{Ca}{2} K
and H lines and, in more extreme cases, infilling and emission in the
hydrogen lines.  We continue the use of the spectral classification 
notation we introduced in paper I to indicate increasing levels of 
chromospheric activity
evidenced by these emission features: (k), k, ke and kee.  In the
first two cases, infilling is seen only in the K and H lines; in the
last two, infilling and emission can be seen in the Balmer lines as
well, especially at H$\beta$.  Because chromospheric emission is time
variable, these activity classes are necessarily time dependent.  We have
therefore noted the observation date in the notes to table 2 for those
stars that have been designated either ``ke'' or ``kee''.

\begin{deluxetable}{rrlccrcrrcrccl}
\tabletypesize{\scriptsize}
\tablecaption{Spectral Types, Basic Physical Parameters and Chromospheric Indices}
\tablehead{\colhead{HIP} & \colhead{HD} & \colhead{SpT} & \colhead{N1} &
  \colhead{$T_{\rm eff}$} & \colhead{$\log g$} & \colhead{$\xi_t$} &
  \colhead{$[M/H]$} & \colhead{N2} & \colhead{S$_{MW}$} &
  \colhead{$\log(R^{\prime}_{\rm HK})$} & \colhead{AC} & \colhead{N3} & \colhead{Obs}} 
\startdata
    57 & 224789\phantom{AB}   & K1 V          &\nodata &  4999 & 4.48 & 1.0 & -0.17 &\nodata& 0.377 & -4.568 &  A &\nodata & CTIO \\
   194 & 225003\phantom{AB}   & F1 V          &\nodata &  7043 & 4.07 & 2.0 & -0.12 &\nodata& \nodata & \nodata & \nodata &\nodata & CTIO \\
   296 & 225118\phantom{AB}   & G8.5 V        &\nodata &  5420 & 4.43 & 1.0 &  0.14 &\nodata& 0.325 & -4.570 &  A &\nodata & SO \\
   436 & 55\phantom{AB}       & K4.5 V        &\nodata &\nodata&\nodata &\nodata &\nodata &\nodata& 0.338 & -4.901 &  I &\nodata & CTIO \\
   522 & 142\phantom{AB}      & F7 V          &\nodata &  6257 & 3.99 & 2.7 & -0.18 &\nodata& 0.171 & -4.853 &  I &\nodata & CTIO \\
   560 & 203\phantom{AB}      & F3 Vn         &\nodata &  6715 & 3.85 & 1.1 & -0.19 &   *   & \nodata & \nodata & \nodata &\nodata & SO \\
\enddata
\tablecomments{Table 2 is presented in its entirety in the electronic 
edition of the Astronomical Journal.  A portion is shown here for
guidance regarding its form and content. See also section 6 for notes
on astrophysically interesting stars.  The column headings have the
following meanings: HIP stands for the designation in the {\it Hipparcos}
Catalogue \citep{esa97}; HD for the HD designation; SpT for the spectral 
type---note that some long spectral types are contained in the notes. An
asterisk in column ``N1'' refers to a note on the spectral type.  Headings
$T{\rm eff}$, $\log g$, $\xi_t$ and [M/H] refer to the basic physical
parameters---the effective temperature (kelvins), the logarithm of the
surface gravity (cm s$^{-2}$), the microturbulent velocity in kilometers
per second, and the overall metal abundance on a logarithmic scale where
[M/H] = 0.0 refers to solar metallicity. 
An asterisk in column ``N2'' refers to a note on the fitting process; a
K in this column indicates that the method used to determine the basic
physical parameters was that adopted for the G and K giants.  $S_{MW}$
is the chromospheric activity index on the Mount Wilson system; 
$\log R^{\prime}_{\rm HK}$, a measure of the chromospheric flux in the
\ion{Ca}{2} K and H lines; ``AC'' indicates the chromospheric activity
class---(VI) very inactive; (I) inactive; (A) active; (VA) very active.
Column ``Obs'' indicates the source of the spectrum---(CTIO) the 1.5~m
Cassegrain spectrograph at Cerro Tololo Interamerican Observatory and
(SO) the Cassegrain spectrograph on the Bok 2.3~m telescope of Steward
Observatory. An asterisk in column ``N3'' refers to a note on 
chromospheric activity.} 
\end{deluxetable}

Spectral types for the stars analyzed in this paper can be found in
Table 2.  Figure 1 shows a classical HR diagram ($M_V$ versus spectral
type) for the stars in this paper and paper I.  One of the reasons why
spectral types are valuable to this project is that they enable us to
refine the stellar census in the solar neighborhood.  Note the stars
in Figure 1 which scatter below the main sequence (white dwarfs are
not plotted in this diagram).  These stars, without exception, have
{\it Hipparcos} \citep{esa97}
parallaxes which place them within 40pc, but which have large parallax
errors, mostly due to binarity.  These stars are indicated in the 
notes to table 2.  Notice as 
well in Fig 1 a group of ``dwarf
stars'' which lie above the main sequence (too far, in most cases, to
be explained by binarity) and which overlap the subgiant branch.
These are an interesting and diverse group of stars.  Most (HIP 3185, 14086,
18432, 46404 and 110618) are high-velocity,
metal-weak stars (for which the luminosity class is notoriously
difficult to determine).  All of these stars have V components of
their heliocentric space velocities $< -50$ km s$^{-1}$ and thus are 
probably thick disk or halo stars, and are most likely evolved. Two others are 
RS~CVn binaries (HIP 16846 and HIP~84586 = HD~155555).  
HD~155555 is probably a pre-main-sequence binary.  Finally, one other 
star, HIP~117815, is a W~UMa eclipsing binary.

\begin{figure}
\figurenum{1}
\epsscale{.9}
\plotone{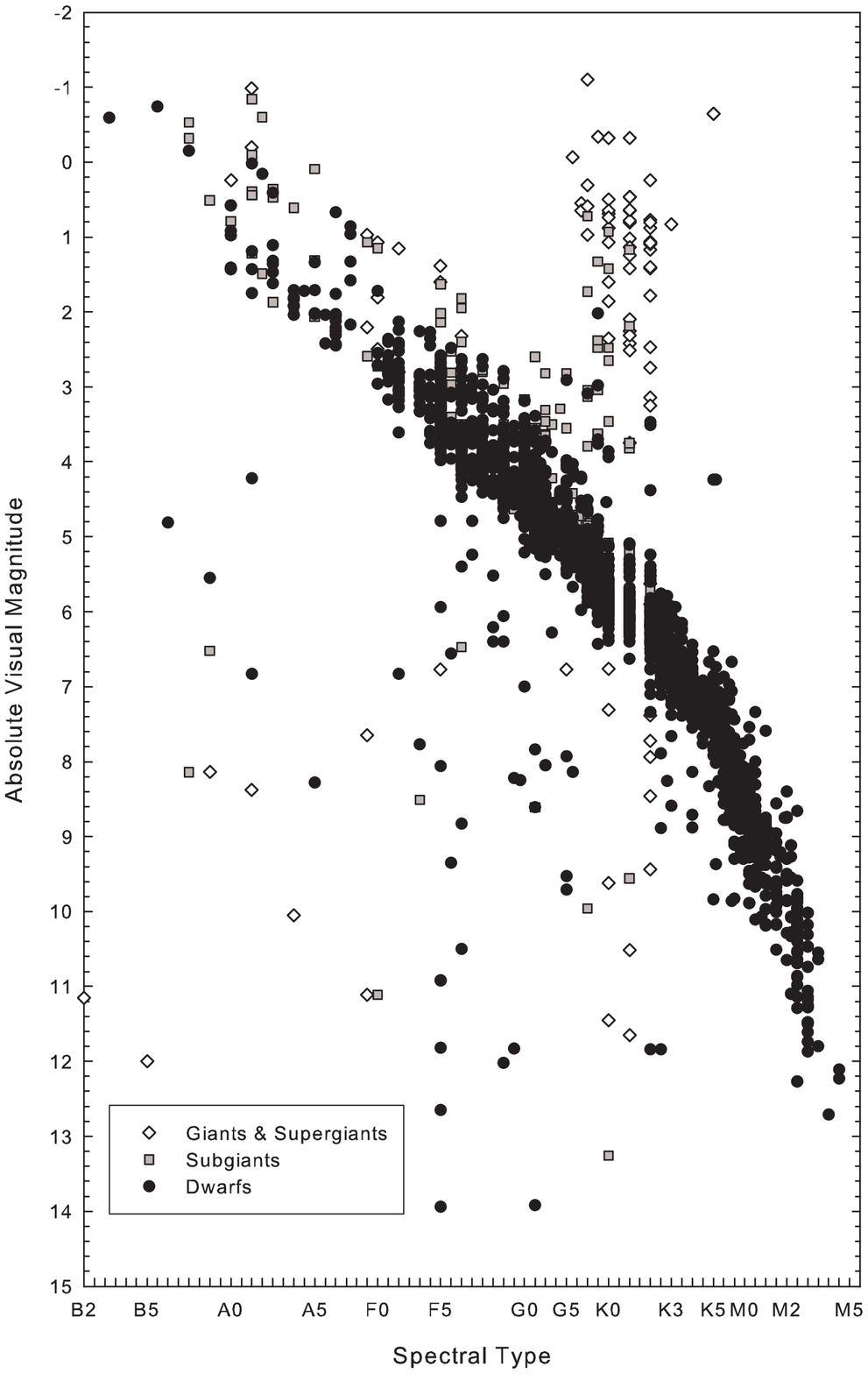}
\epsscale{1}
\caption{An observational HR diagram formed from the spectral types of
the 1670 stars reported in this paper and the 665 stars reported in 
paper I.  The absolute magnitudes, $M_v$, are
calculated using {\it Hipparcos} (ESA 1997) parallaxes.  The stars
scattering below the main sequence in this figure all have large parallax
errors and are listed in the notes to Table 2.  These stars are all 
evidently more distant than 40pc. }
\end{figure}

Spectral types also provide beginning values for our determination of
the basic physical parameters and provide a check on the derived
physical parameters.  But, most importantly, spectral types place a
star within the context of a broad population of stars, and enable us
to pick out peculiar and astrophysically interesting stars.

\section{Basic Physical Parameters}

An important goal of this project is to determine the basic physical
parameters---the effective temperature, the surface gravity and the
overall metallicity---for as many of our program stars as possible.

\subsection{The Determination of the Basic Physical Parameters}

We described in paper I the technique that we use to determine the
basic physical parameters for our program stars, and we refer the
reader to that paper for the details.  In summary, we determine the
basic physical parameters by carrying out a simultaneous fit, using a
variant of the multi-dimensional downhill simplex method
\citep{press92}, between 1) the observed spectrum and a library of
synthetic spectra and 2) observed fluxes from Str\"omgren $uvby$ and
Johnson/Cousins $VRI$ 
photometry \citep[taken from ][]{mermilliod97} and theoretical fluxes 
based on \citet{kurucz93} ATLAS9
stellar atmosphere models (computed without convective overshoot).
The library of synthetic spectra was calculated
with the stellar spectral synthesis code SPECTRUM
\citep{gray94}\footnote{see 
http://www.phys.appstate.edu/spectrum/spectrum.html} using ATLAS9
models.  A graphics program, xfit21, written by one of us (ROG) is
used to set the initial parameters which are then polished by the 
simplex engine.  This graphical front end allows us to confirm that 
the simplex engine has indeed found the optimal global solution, and 
it also allows us to tweak that solution if necessary.  The program
xfit21 allows the user to apply rotational broadening when necessary,
and also to deredden the observed fluxes.  However, since all of our
stars are within 40pc, we have assumed the reddening for our stars to
be zero.

As detailed in paper I, for certain spectral types it was necessary to
constrain one or more of the physical parameters in order to achieve
reasonable solutions.  This was especially the case for late-G and K dwarfs
and giants.  For the late-G and K dwarfs, neither our spectra nor the
photometry strongly constrain the surface gravity.  For these stars we
have therefore constrained $\log g$ to the value implied by the {\it
  Hipparcos} parallax and the mass-luminosity relationship
\citep{gorda98}. We have further constrained the microturbulent
velocity $\xi_t$ to be $1.0$km s$^{-1}$.  However, we encountered an
additional difficulty with the SO and CTIO spectra.  Because these
spectra have a shorter spectral range than the 
DSO 3.6\AA\ spectra used in paper I for the late-G and K dwarfs, and
thus span only two Str\"omgren bands ($v$ and $b$) instead of
three, only a linear instead of a second order spline could be used to
apply the photometric correction.  In addition, a significant number 
of the SO spectra were obtained at
quite high air masses, and thus the flux calibration for those spectra
is quite unreliable.  Because the accuracy of the flux calibration of 
the observed spectrum is of some importance in obtaining a reliable
fit to the basic physical parameters for the late-type stars, we adopted the
following procedure to overcome this difficulty, made possible by the 
xfit21 program:  Having first constrained the gravity to the value 
implied by the mass-luminosity relation and the {\it Hipparcos}
parallax, and fixed $\xi_t$ at 1.0~km~s$^{-1}$, we set
[M/H]$ = 0.0$ and then adjusted the effective temperature to
minimize the residuals between the photometric and theoretical fluxes.
We then visually adjusted [M/H] to match the line strengths in the
observed and synthetic spectra, and then iterated until the match was
satisfactory.  At this point, if the photometric correction of the
observed spectral fluxes was satisfactory, the residuals between the 
observed and synthetic spectrum would be flat.  If, however, the photometric
correction was not satisfactory, these
residuals might display a slope or a shallow curve as a function of the 
wavelength.  If such was the case, the synthetic spectrum was used as
a template to correct the fluxes in the observed spectrum.  The
simplex engine was then allowed to polish the final solution.  While
this correction procedure could be iterated, in practice we found one
application of this procedure gave a satisfactory fit to the basic
physical parameters.  Illustrations of typical simplex solutions 
may be found in paper I.

\begin{figure}
\figurenum{2}
\plotone{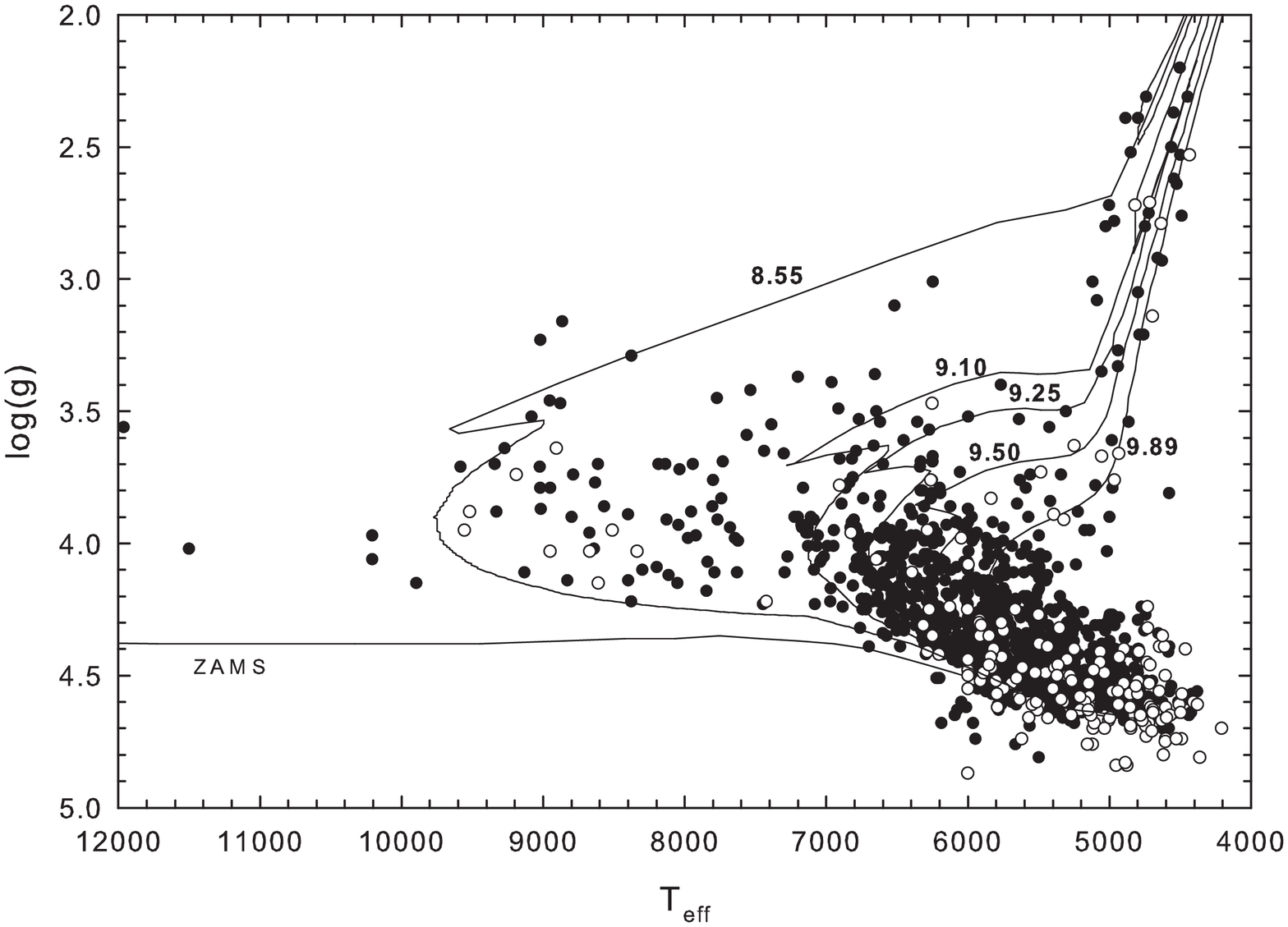}
\caption{An astrophysical HR diagram based on results reported in this
paper and paper I.  The isochrones, included for illustrative
purposes, are from \citet{lejeune01}.  The isochrones are labeled with
the logarithm of the age (8.55 = 350 million years, 9.10 = 1.3 Gyr,
9.25 = 1.8 Gyr, 9.50 = 3.5 Gyr and 9.89 = 7.8 Gyr).  The filled
circles are for stars with [M/H] $> -0.40$, open circles for stars
with [M/H] $< -0.40$.}
\end{figure}

Effective temperature determinations for the stars in common between
SO and CTIO give us an excellent opportunity to test whether this modified
method introduces any systematic error into the determination of the
basic physical parameters.  The reason for this is that most of these
stars were obtained at high airmass from SO, but were observed nearly
overhead from CTIO.  For the 56 dwarf stars in common between SO and
CTIO with basic physical parameters, we find the following systematic
difference in the effective temperatures:
\begin{displaymath}
{\rm T_{CTIO} - T_{SO}} = 0{\rm K} \pm 40{\rm K}
\end{displaymath}
where the uncertainty is for a single point.  This also gives a good
estimate for the internal error in the effective temperatures; the
above comparison suggests that the internal precision in the
temperature determinations is on the order of $\pm 30$K.  For those dwarfs 
with $V-K$ photometry, we may also
compare our temperature scale with that of the Infrared Flux method
\citep{blackwell94} which is essentially independent of theoretical
models (see also paper I).
We find,
\begin{displaymath}
\begin{array}{llrcrr}
{\rm T_{CTIO} - T_{IRFM}} & = & -4{\rm K} & \pm & 16{\rm K} & {\rm (29\;  stars)} \\ 
{\rm T_{SO} - T_{IRFM}} & = & 32{\rm K} & \pm & 29{\rm K} &  {\rm (12\; stars)} \\
\end{array}
\end{displaymath}
and, from paper I,
\begin{displaymath}
\begin{array}{llrcrr}
{\rm T_{DSO} - T_{IRFM}} & = & 28{\rm K} & \pm & 17{\rm K} & {\rm (50\; stars)} \\
\end{array}
\end{displaymath}
where the error displayed in all three cases is the error in the 
determination of the zeropoint difference, and not the scatter around
the IRFM relationship.  Combining the CTIO and the SO datasets, the
scatter around the IRFM relation is $\pm 90$K, some of which is due to
errors in the $V-K$ photometry (a change of 0.04 in $V-K$ at $V-K = 2.00$
results in a temperature change of 50K).  A plot of the
residuals with respect to the IRFM relation shows no trend with 
effective temperature over the range 4600 -- 8000K.   All 
three datasets show acceptably small and 
statistically indistinguishable zeropoint differences with the IRFM
method.

There are now a number of spectroscopic studies in the literature dealing with
nearby stars with which we may profitably compare our temperature
scale. \citet{prieto04} used high resolution spectroscopy to study the
stars with $M_V < 6.5$ within 14.5~pc of the sun.  They base their 
temperatures on the calibrations of 
\citet{alonso96,alonso99} who in turn derived their calibrations of
\bv\ and Str\"omgren photometry from the IRFM method.  Our temperatures 
compare well with theirs, except for a slight trend in the residuals
which amounts to our temperatures being about 50K cooler than theirs
at 5000K, and 120K hotter at 6000K.  Once this trend is removed, the
temperatures agree to within $\pm 70$K.  The origin of this 
trend is not clear.  \citet{luck05} have studied the stars with $M_V <
7.5$ within 15~pc of the sun, but, unlike Allende Prieto et al., their
atmospheric parameters were determined by requiring the abundance of 
iron to be independent of the excitation energy simultaneously with
requiring that \ion{Fe}{1} and \ion{Fe}{2} give identical iron
abundances.  Our temperatures compare quite favorably with theirs with
a zeropoint offset of 89K (our temperatures are cooler) and a scatter
of $\pm 108$K, but no trend in the residuals.  Finally,
\citet{valenti05} have studied 1040 nearby F, G and K-type stars which
have been observed in the Keck, Lick and AAT planet search programs.
They have used high resolution spectra and a pipeline multivariate method to 
determine their basic physical parameters.  Our temperature scales are
virtually identical for the F and G-type stars, but cooler than 
$T_{\rm eff} = 5200$K, their temperatures begin to systematically
deviate from ours (our temperatures are cooler) with the mean deviation
approaching 100K at 4800K.  For stars with effective temperatures
greater than 5200K, the scatter in the comparison of our temperatures
with those of Valenti \& Fischer is only $\pm 81$K, with a zeropoint
difference of 14K (our temperatures are cooler).  

In paper I we adjusted the zeropoints of our [M/H] determinations
using well studied stars in the \citet{cayrel01} [Fe/H] catalog, and
we have done the same in this paper.
Table 3 lists these zeropoint corrections, which are generally quite
small and similar in magnitude to those found in paper I with the
exception of the K giants (see below).

\begin{deluxetable}{lll}
\tablewidth{0pt}
\tablenum{3}
\tablecaption{Zeropoint Corrections to [M/H] scale}
\tablehead{\colhead{Stellar Type} & \colhead{CTIO} & \colhead{SO}}
\startdata
F and G dwarfs & +0.02 & +0.04 \\
G and K dwarfs & +0.07 & +0.07 \\
G and K giants & +0.08 & +0.06 \\
\enddata
\end{deluxetable}

It is of interest to compare the [M/H] scales for the CTIO and SO
spectra.  We find, for the 56 stars in common
\begin{displaymath}
{\rm [M/H]_{CTIO} - [M/H]_{SO}} = 0.00 \pm 0.08
\end{displaymath}
where the quoted error is for a single point.
Of course, the negligible zeropoint difference is not surprising, as
these metallicities include the zeropoint corrections of table 3, but
the quoted error can be used to estimate the internal precision
of the [M/H] determinations.  This is evidently on the order of 
$\pm 0.06$~dex.  Once the zeropoints have been applied (the [M/H]
values in Table 2 include these zeropoint adjustments), the agreement
with mean values in the [Fe/H] catalog is excellent, with a scatter
of $\pm 0.09$~dex.  Comparing the [M/H] values in Table 2 with 
the \citet{valenti05} sample, however, we find an additional zeropoint 
difference of 0.07~dex (our values are more negative), but a scatter of only 
$\pm 0.09$~dex. 

The method of determining the basic physical parameters for late G and
K giants was essentially the same as described in paper I, although
our improved method of applying a flux template at the midpoint of the
fitting process (described above) appears to have significantly
improved the quality of the fit for the K giants.  Indeed, the
zeropoint correction for the K giant [M/H] scale (see table 3) is considerably 
smaller than reported in paper I, and is now comparable to the
corrections for the dwarf stars.

Figure 2 is a theoretical HR diagram based on results from paper I and
this paper.  Note the excellent match between the position of the
giant branch and the isochrones, as well as an indication of an
``edge'' in the distribution of stars corresponding to the 1.8 Gyr
isochrone.  We will analyze this diagram in paper III for the complete
sample.

\section{Chromospheric Emission}

All of the spectra obtained for this project include the \ion{Ca}{2} K
and H lines and thus can be used to obtain measures of the
chromospheric emission via emission reversals in the cores of these
very strong lines.  This is an important measurement, as chromospheric
emission can be an indication of the age of a star and/or its binary
status.  An age determination can be important for exoplanet searches,
especially for both the nulling-interferometer and coronographic
designs of the {\it Terrestrial Planet Finder}, because strong zodiacal
light, scattered from a remnant protoplanetary disk, could mask weak
planetary signals.  

Paper I described in detail how we measure the chromospheric emission
in our program stars.  We follow the practice of the Mount Wilson
chromospheric activity program \citep{baliunas95} in measuring the
fluxes in four bands; two of these bands, C$_1$ and C$_2$, are
continuum bands located just shortward and longward of the K and H
lines, and the K and H bands are centered on the cores of the K and H
lines (see Fig 9 in paper I).  Our K and H bands are 4\AA\ wide, in
contrast to the Mount Wilson 1\AA\ bandpasses because of the lower
resolution of our spectra.  The chromospheric emission index is 
calculated, like the Mount Wilson index, with the equation
\begin{displaymath}
S = 5\frac{K + H}{C_1 + C_2}
\end{displaymath}
Because the CTIO and SO spectra have different resolutions, there is a
different instrumental system for each.  The ultimate goal is to
transform these instrumental systems on to the system of the Mount Wilson
project, but to ensure homogeneity in the chromospheric indices
reported in this series of papers, we first transform these two instrumental
systems to the DSO18 instrumental system established in paper I.  The
DSO18 instrumental system ($S_{18}$) is then transformed to the Mount Wilson
system ($S_{\rm MW}$) using observations of a number of chromospheric activity
``standards''---stars well observed by the Mount Wilson group and
which, at least in their database ending in 1995, show no significant
long term secular trends (see table 5 in paper I).  This
transformation from $S_{18}$ to $S_{\rm MW}$ was derived in paper I.

To derive the transformations between $S_{\rm CTIO}$ and $S_{18}$ and
between $S_{\rm SO}$ and $S_{18}$ we observed as many of the
chromospheric ``standard'' stars mentioned in the above paragraph as
possible from CTIO and SO.  The resulting transformations are linear:
\begin{displaymath}
\begin{array}{lcll}
S_{\rm 18} & = & 0.949S_{\rm CTIO} - 0.032 & \sigma = 0.012 \\
S_{\rm 18} & = & 0.904S_{\rm SO} + 0.002   & \sigma = 0.011 \\
\end{array}
\end{displaymath}
and are illustrated in Figure 3.

\begin{figure}
\figurenum{3}
\plotone{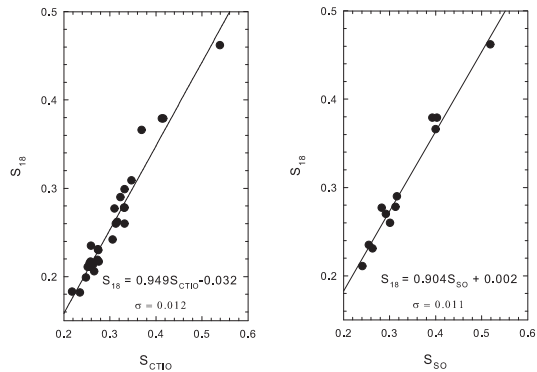}
\caption{Transformations from the CTIO and SO chromospheric emission 
instrumental systems to the DSO18 instrumental system of paper I.}
\end{figure}  

As explained in paper I, we have used the procedure of \citet{noyes84}
to derive from $S_{\rm MW}$ the parameter $\log R^{\prime}_{\rm HK}$
which is a measure of the chromospheric flux in the cores of the K and
H lines.  This parameter may be used to classify stars into the
chromospheric activity categories ``very inactive'' (VI), ``inactive''
(I), ``active'' (A) and ``very active'' (VA) (see Fig 4).  This
parameter has been
tablulated for stars in our sample with spectral types between F5 and
M0 (see table 2).  However, the reader should bear in mind that the 
transformation from $S_{\rm MW}$ to $\log R^{\prime}_{\rm HK}$ becomes
increasingly uncertain for stars with $\bv > 1.2$.  In addition, this
transformation is not well defined for $\bv < 0.5$.  Thus, even though
$\log R^{\prime}_{\rm HK}$ is tabulated in table 2 for stars outside
of this range, these values and the corresponding activity
classifications should be treated with some caution.  The 
distribution of this parameter with respect to metallicity is discussed in
\S~7.1, and that discussion is limited to stars with $0.5 < \bv
< 1.2$.

\begin{figure}
\figurenum{4}
\plotone{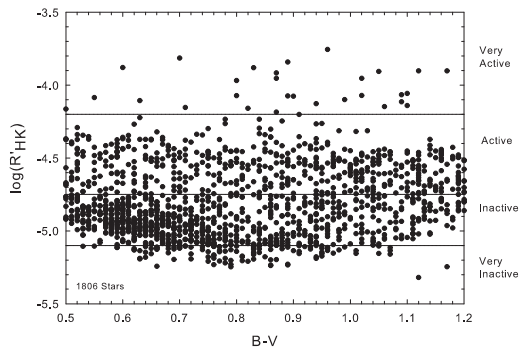}
\caption{Plot of the chromospheric flux parameter $\log
  R^{\prime}_{HK}$ vs. the $B-V$ color.  This diagram allows the
  classification of stars into the chromospheric activity categories
  ``very inactive'', ``inactive'', ``active'' and ``very active''.}
\end{figure} 

A brief note on the relationship between our spectral-classification
notation for chromospheric activity (see \S~3) and the classification
into the activity classes VI, I, A and VA afforded by the 
$\log R^{\prime}_{\rm HK}$ parameter is probably appropriate here.
These two classification
systems are generally in good agreement.  For instance, nearly all of
the ``ke'' and ``kee'' stars are either in the ``A'' (Active) or
``VA'' (Very Active) classes.  However, the two systems
are not redundant.  The $\log R^{\prime}_{\rm HK}$ classification uses
information external to the spectrum, in particular, the \bv\ index and
a model of the photospheric flux in the K-line.  It is also
not defined for late K and early M dwarfs, nor does it contain any
information about emission in the hydrogen lines.  The spectral
classification notation, on the other hand, is self-sufficient; 
it may be used for M dwarfs and it also contains information on 
Balmer-line emission, which is not necessarily well correlated with 
emission in \ion{Ca}{2} K \citep[see ][]{thatcher93}.

\section{Notes on Astrophysically Interesting Stars}

\noindent HIP~3961 = HD~5028:  We were somewhat surprised to find 
that this metal-weak F6 star turns out to be in the chromospherically 
very active category.  However, visual inspection of the spectrum
verifies that the \ion{Ca}{2} K \& H lines are both shallow.  It turns
out that this star is both an X-ray source \citep{haberl00}  and a 
far-UV source \citep{bowyer95}.

\noindent HIP~2235 = HD~2454: An F5 dwarf with an over abundance of Strontium.
Ba II 4554 also appears enhanced---see \citet{tomkin89}.

\noindent HIP~16846 = HD~22468 = V711~Tau:  K2: Vn k. 
This well-known RS~CVn variable 
shows strong emission in \ion{Ca}{2} K \& H with infilling in H$\beta$.  The 
spectral lines appear broad.

\noindent HIP~29804 = HD~43848:  This K2 subgiant shows a strong Swan 
band at $\lambda$4737. 

\noindent HIP~30476 = HD~45289: From our spectral type, basic physical
parameters and chromospheric activity measurements, this star appears
to be a very close solar twin (see \S~7.2).

\noindent HIP~59750 = HD~106516: Both the SO and the CTIO spectra agree that
this metal-weak F9 dwarf is a chromospherically active star.  The CTIO 
spectrum, obtained on December 14, 2002 gives 
$\log(R^{\prime}_{HK}) = -4.158$ (very active) and the SO spectrum,
obtained on April 9, 2001 gives $\log(R^{\prime}_{HK}) = -4.410$ 
(active).  This star is an X-ray source, and {\it may} have been the source
of the January 13, 1993 gamma-ray burst---see \citet{shibata97}.

\noindent HIP~64478 = HD~114630: G0 Vp k.  The entire spectrum of this 
chromospherically active star appears ``veiled''.  The
resonance/low-excitation lines (such as \ion{Ca}{1} 4226, \ion{Fe}{1} 4046,
etc.) are particularly weak, the cores of \ion{Ca}{2} K \& H are
shallow, and H$\beta$ appears slightly filled in.  See a related
discussion in \S~7.1.

\noindent HIP~71908 = HD~128898: This well-known SrCrEu Ap star is also 
characterized by a broad \ion{Ca}{2} K-line.

\noindent HIP~76550: This star shows peculiar morphology in the 
$\lambda$4780 MgH band (violet side weak), seen
in both the CTIO and SO spectra.

\noindent HIP~96635 = HD~185181: This is a chromospherically active 
subgiant K2
star and thus a possible PMS star.  Note that \citet{koen02} have found
that this star is a variable from Hipparcos photometry, but were
unable to determine the type of variability. 

\noindent HIP~98470 = HD~189245: This chromospherically very active late
F-type star is an extreme ultraviolet source, a variable star and a 
rapid rotator (86 km s$^{-1}$).

\noindent Also see notes on specific stars in \S~7.1, and the 
notes to table 2.

\section{Results and Discussion}

\subsection{Stellar Activity and [M/H]}

In paper I we demonstrated that the chromospheric
emission parameter $\log R^{\prime}_{\rm HK}$ has a bimodal
distribution (see Figure 12 of that paper).  This is a manifestation
of the well-known Vaughan-Preston Gap \citep{vaughan80}.  However, it 
turns out that this bimodality is a strong function of metallicity
(see Figure 5); for stars with [M/H] $> -0.20$, the
distribution is strongly bimodal (Fig 5c), but the
distribution is strictly single-peaked for stars of lower metallicity
(Fig 5b).   This is perhaps not too surprising, as metal-weak stars
will, on the average, be older than metal-rich stars, and thus we
should expect a smaller number of active metal-weak stars, but two things
are remarkable: 1) the fact that there persists a tail of quite active
stars even at quite low metallicities ([M/H] $\approx -0.50$) and 2) the
sudden change to a single-peaked distribution at [M/H]$ = -0.20$.  We 
also note 
that the low-metallicity peak is somewhat shifted towards less
negative values of $\log R^{\prime}_{\rm HK}$ (i.e. towards higher
activity levels).

\begin{figure}
\figurenum{5}
\plotone{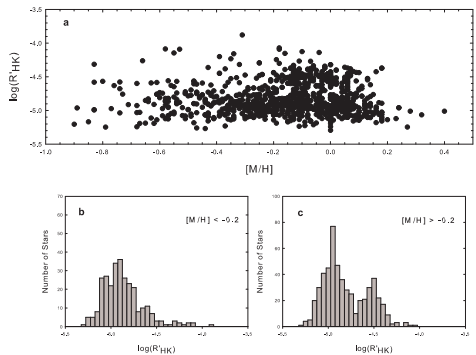}
\caption{The distribution of the chromospheric activity parameter
$\log R^{\prime}_{\rm HK}$ for dwarf F, G and early K-stars as a
  function of metallicity.  In (a), activity increases along the
  vertical axis, in (b) and (c) to the right.  Panels (b) and (c) show
  histograms of the distribution of points in panel (a).  Note that
  for stars with [M/H]$ > -0.2$ the distribution is strongly bimodal,
  but it is single-peaked for [M/H]$ < -0.2$.  Also note that a tail
  of very active dwarf stars persists in the metal-weak distribution.
  Both of these points are discussed in \S~7.1}
\end{figure}  

Let us consider this last point before we discuss points (1) and (2).
The slight shift in the low metallicity peak towards higher activity
levels can be understood as a metallicity effect on the fluxes in the
four bands used in the calculation of the chromospheric activity
parameters $S_{\rm MW}$ and $\log R^{\prime}_{\rm HK}$ (see \S~5).  To
illustrate this, we have calculated both of these parameters using
synthetic spectra with $T_{\rm eff} = 5000$K and $\log g = 4.5$ with
[M/H] ranging from $-1.0$ to 0.5 (see table 4).  Note that the activity
class of these synthetic spectra (which do not include chromospheres)
changes from very inactive (VI) for the metal-rich spectra to inactive
(I) for the metal-weak spectra.  This suggests a need to update the
procedure of \citet{noyes84} to calculate $\log R^{\prime}_{\rm HK}$
by including a correction factor for metallicity, but it 
also helps to explain the shift in
the low-metallicity peak towards higher activity levels and the fact that
only a few of the metal-weak stars in table 2 have activity classes
of VI.  However, it does not help to explain the presence of an
extended tail of very active stars visible in Fig 5b, which brings us
back to point (1) above. 

\begin{deluxetable}{rlllr}
\tablewidth{0pt}
\tablenum{4}
\tablecaption{Activity Class and Metallicity}
\tablehead{\colhead{[M/H]} & \colhead{\bv} & \colhead{$S_{\rm MW}$} & 
\colhead{$\log R^{\prime}_{\rm HK}$} & \colhead{AC}}
\startdata 
 0.50 & 0.99 & 0.125 & -5.279 & VI \\
 0.00 & 0.92 & 0.125 & -5.245 & VI \\
-0.50 & 0.87 & 0.134 & -5.185 & VI \\
-1.00 & 0.84 & 0.159 & -5.054 & I \\ 
\enddata
\tablecomments{Note: VI represents the ``Very Inactive'' and I the
  ``Inactive'' chromospheric activity classes (see text).}
\end{deluxetable}

The nature of these ``low-metallicity'' chro\-mo\-spher\-ically-active 
stars may be explored by considering some examples.  

HD~9054 = HIP~6856 = CC~Phe is a chromospherically active K2 dwarf 
(K2+ V k) with $\log R^{\prime}_{\rm HK} = -4.263$ (on the boundary
between the Active and Very Active categories).  We have calculated [M/H]$ =
-0.66$.  This star is a strong 
X-ray source, and \citet{torres00} have found that this star is a
member of a very young nearby association,
HorA, in the vicinity of the active star ER Eri.  In this context the
metal-weak nature of this star is very difficult to understand.

HD~146464 = HIP~79958 = V371~Nor is a very active K3 dwarf (K3 V ke)
which is not well studied.  Its position coincides closely with 
1RXS J161915.7-553023, an X-ray source.  Kinematically, this star
appears to be thin disk star ($U, V, W = 3,1,1$~km s$^{-1}$)---all
kinematics quoted in this paper are heliocentric space velocities
in a right-handed coordinate system with U pointing toward the
galactic center and are taken from \citet{nordstrom04}---but we 
have calculated [M/H]$ = -0.55$.

Two BY Draconis variables from paper I are also found to have low
metallicities.  These stars are:

HD~45088 = HIP~30630 = OU~Gem, which has a high eccentricity orbit ($e
= 0.15$) and kinematics of a thin disk star: 
$U, V, W = +9, -4, -11$~km s$^{-1}$, but we found [M/H]$ = -0.83$.
\citet{soderblom93} list this star as a possible member of the UMa
Group (age $\approx 0.3$~Gyr).

HD~218738 = HIP~114379 = KZ~And, which also has kinematics of a thin
disk star: $U, V, W = -8, -10, -3$~km s$^{-1}$.   We found [M/H]$ = -0.53$.

These four stars are surprisingly metal-weak for active, low-velocity
stars.  Our metallicities for three of these four stars are in 
good agreement with those of the Geneva-Copenhagen group \citep{nordstrom04}
who base their metallicities on a calibration of the Str\"omgren $m_1$
index.  \citet{nordstrom04} do not give a metallicity for HD~146464.  
Other less extreme examples may be found by examining table 2.
This observation, that many chromospherically active K dwarfs have ``low''
metal abundances is not new.  For instance, \citet{rocha98} examined
the $\log R^{\prime}_{\rm HK}$ index for a sample of G and K dwarfs
and found that many of the most active stars had low values of the 
Str\"omgren $m_1$ index, which measures line blanketing in the violet
part of the spectrum.  \citet{giampapa79} observed solar active and
quiescent regions with the Str\"omgren $uvby$ system and found that
the solar active regions are up to 35\% more metal deficient than the
quiescent regions, based on the $m_1$ index.  \citet{favata97} also
demonstrated the depression of the $m_1$ index for active K-stars, but
obtained red spectra for a number of these stars and showed that these
spectra gave normal (roughly solar) abundances.  They suggested that the
depression of the $m_1$ index is due to emission or infilling of the
H$\delta$ line which lies in the Str\"omgren $v$-band.  However, a
careful examination of the spectra of our four low-metallicity active
K dwarfs 
does not show emission, or even infilling of the H$\delta$ line (at
least in ratio with nearby metallic lines), with
the possible exception of HD~146464 which may
show a slight infilling of the H$\delta$ line.  On the other hand, all 
of these
stars show a marked weakening of the line spectrum in the blue-violet
region (up to at least 4400\AA) relative to solar-abundance
chromospherically inactive stars of the same spectral types, easily
apparent even on visual inspection (we did not use the MK K-dwarf standards
for this comparison, as some of these standards are metal weak, while
others are chromospherically active, and may show some veiling
themselves).  The veiling seems most pronounced 
in the \ion{Ca}{1} 4226 line and in the vicinity of the CN 0,1 band
with bandhead at 4216\AA.  \citet{basri89}, using high resolution 
spectra, have shown that the equivalent width of metallic lines in the 
blue-violet region can be reduced in highly
active stars, which correlates well with what we see in our
blue-violet spectra.  Whether this infilling of the metallic lines is
due to line emission or continuum emission is not clear at this point, 
but this phenomenon does seem to be the best explanation for the
observed metallicity effect.

We note, however, that there are a few very active K dwarfs in
our sample which do not seem to show a pronounced metallicity effect.  
For instance, HD~26354 (AG~Dor), HD~111038 (LO~Mus) and HD~220140
(V368 Cep) have K-line emission and  
$\log R^{\prime}_{\rm HK}$ indices which place them in the very active 
category, but have, according to our determinations, metallicities
that are only slightly sub-solar.  
Both HD~26354 and HD~220140 have been classified as RS~CVn binaries.  
LO~Mus is not well studied, but appears to be a BY Draconis star 
\citep{kazarovets97}.
Other less extreme examples can be found in table 2.  HD~26354 does
not show any veiling, and even has a slightly stronger CN band than 
normal K2 solar-metallicity dwarfs (its gravity, however, is
consistent with it being a dwarf and not a subgiant).  HD~220140 also
has a gravity consistent with a dwarf classification.   HD~111038 shows 
marginal veiling, and
HD~220140 shows some veiling, especially in the far violet
(just longwards of the K \& H lines) and in \ion{Ca}{1} 4226, which
also appears slightly weak in HD~111038.  To add to the mystery, however,
the Geneva-Copenhagen group \citep{nordstrom04} find, from their
calibration of the Str\"omgren $m_1$ index, that these three stars
are quite metal-weak.  

It is difficult to understand why some of these very active K
dwarfs show a pronounced metallicity effect and veiling and others do
not.  Both groups contain active binary stars and both contain objects 
which are EUVE and X-ray sources.  We speculate, however, that the 
chromospheres of these two groups may differ significantly.  For instance, the
``veiling'' effect that is more prevalent in the metallicity-effect
stars is largely seen in resonance or low-excitation lines of neutral
metals in the violet region of the spectrum, such as \ion{Ca}{1} 4226.
The cores of such lines are formed in the temperature minimum
region whereas the emission reversals in the \ion{Ca}{2} K \& H lines
are formed higher in the chromosphere in G and K dwarfs.  This may
mean that the temperature structure and/or the density in the
temperature minimum region is different in the metallicity-effect
active K-dwarfs.  Additional observations, to quantify and to determine the
nature of the infilling of the metallic-line spectrum in the
metallicity-effect objects, will be required to resolve this question.

We now consider point 2.  The sudden change from bimodality to
single-peaked behavior at [M/H]$ = -0.2$
becomes even more pronounced when the metallicity effects discussed in
the above paragraphs are taken into account, because many (although
not all) of the active ``low metallicity'' dwarfs in the high activity
tail in the [M/H] $ < -0.2$ distribution really belong at [M/H] $ >
-0.2$.  Two possible exceptions are HD~9770, an active triple
(quintuple?) system, which has kinematics typical of an old disk star 
\citep[see ][]{eggen62, watson01} and BF~Lyn, a BY~Dra variable with
kinematics ($U,V,W = -22,-51,-28$~km s$^{-1}$) suggestive of a thick disk star.
This sharp transition from bimodality to single-peaked behavior at 
[M/H] $ = -0.2$ suggests that the cause of this phenomenon is not 
primarily age related,
but rather is associated with some parameter necessary for the 
generation of an active 
chromosphere which is switched off at this divide.  We expect this parameter
has something to do with rotation or, more specifically, differential 
rotation, but we do not have sufficient data to speculate further.

\subsection{Solar Analogues}

In Table 5 we have, for the benefit of users who are interested in
solar analogues and twins and exoplanet searches, extracted from Table
2 all those dwarf stars (a total of 61) which satisfy the following 
requirements: 1)
Spectral types between G0 and G5, 2) $\log g > 4.20$, 3) [M/H] $ >
-0.10$, 4) single or members of wide doubles.  In this table we have
further distinguished stars by those which have spectral types, basic
physical parameters and activity levels which are close (*) and very 
close (**) to those of the sun.  Indeed, those marked ** in the
table can be considered solar twin candidates.  We have also indicated
in this table those which have known exoplanets, and those which are
currently on the Keck, Lick \& AAT Doppler planet-search program,
according to \citet{valenti05}.

\begin{deluxetable}{rrr}
\tablewidth{0pt}
\tablenum{5}
\tablewidth{0pt}
\tablecaption{Solar Analogues}
\tablehead{\colhead{HIP} & \colhead{HD} & \colhead{Notes}} 
\startdata
   699$\ldots$ & 361$\ldots$ & * \\
  1444$\ldots$ & 1388$\ldots$ & D  \\
  1954$\ldots$ & 2071$\ldots$ & *D \\
  3578$\ldots$ & 4392$\ldots$ &   \\
  6455$\ldots$ & 8406$\ldots$ & * \\
\enddata
\tablecomments{Table 5 is presented in its entirety in the electronic 
edition of the Astronomical Journal.  A portion is shown here for
guidance regarding its form and content.  Symbols in the notes column 
have the following
  meanings: *; A solar analogue with spectral type within 1 subclass
  of the sun.  **; A candidate solar twin with spectral type, and
  physical parameters very similar to that of the sun.  D;  A star
  which is already in the Keck, Lick \& AAT Doppler planet-search
  program (see Valenti \& Fischer 2005). EP; A star with
  a known exoplanet.}
\end{deluxetable}

\subsection{Metallicity distribution of the Solar Neighborhood}

Information regarding the star formation and the chemical enrichment
history of the galactic disk can be derived from the metallicity
distribution of the solar neighborhood.  We have used data in table 2
of this paper and table 1 of paper I to plot a histogram of [M/H]
for our sample of stars.  This histogram, based on 1364 stars, appears 
in Figure 6.

\begin{figure}
\figurenum{6}
\plotone{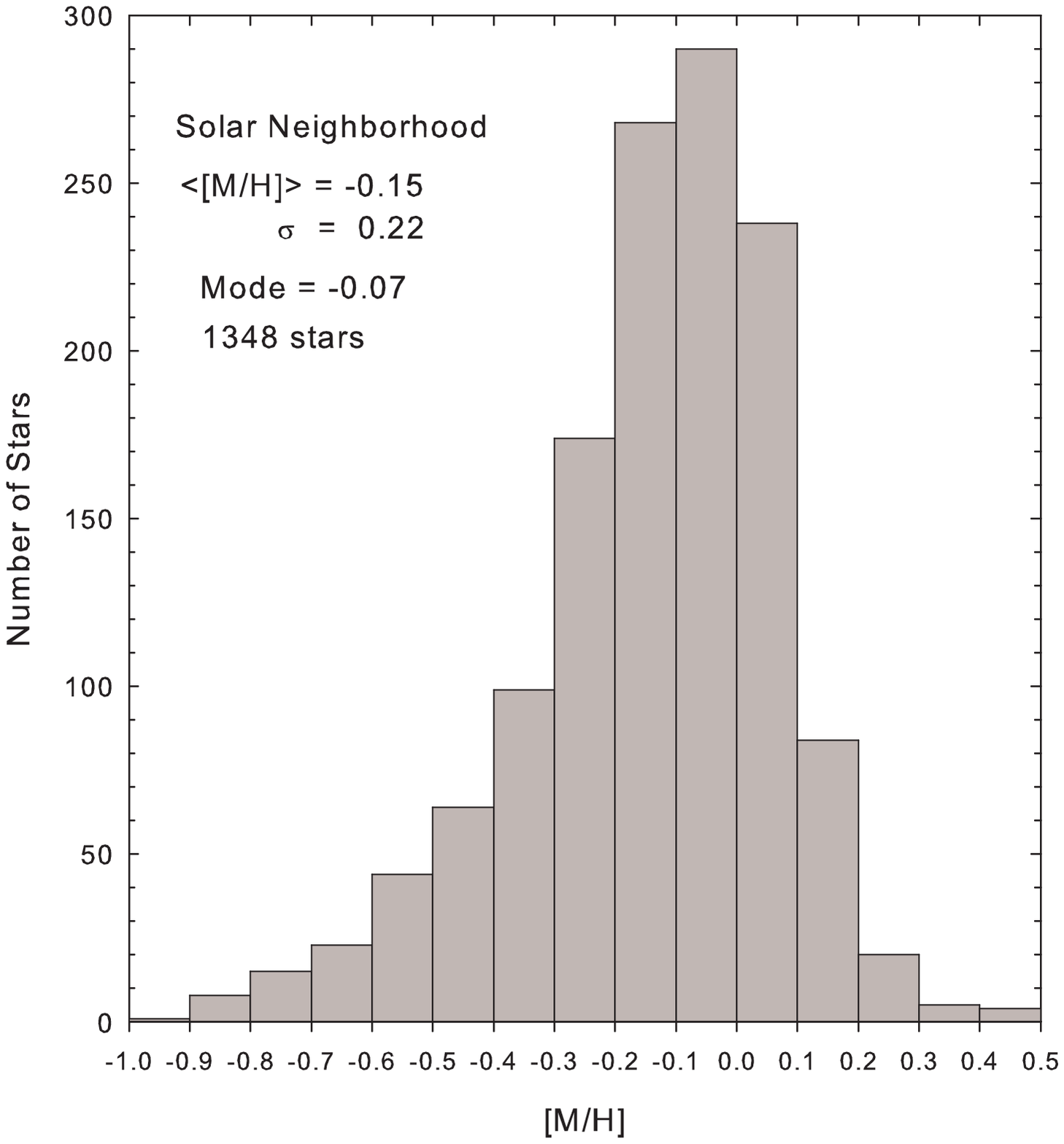}
\caption{The histogram shows the metallicity distribution for stars
  in our sample in the solar neighborhood.  Note the extended
  low-metallicity tail, composed mostly of stars from the thick disk.}
\end{figure}

It is clear that this distribution is basically Gaussian with an
enhanced low metallicity tail, probably due to thick-disk stars, some
``metallicity-effect'' active K-stars (see \S 7.1) and even
a few interloping halo stars (our sample, for instance, includes HD~19445, a
well-studied halo star with [Fe/H]$ \approx -2$).  Taking a straight
mean, we find $\langle$[M/H]$\rangle = -0.15$ with a dispersion of 0.22
dex.  This is in excellent agreement with the recent determination by
the Geneva-Copenhagen survey \citep{nordstrom04} (-0.14, $\sigma =
0.19$) and is also similar to one found for K-type giants by \citet{girardi01}.

However, if one is interested in the metallicity distribution of the
thin-disk population, thick disk and halo stars in the low metallicity 
tail must be removed by reference to kinematics.  We will carry out 
this analysis in paper III where we will publish results for the
remainder of our sample, but it is clear that the result will be close 
to the {\it mode} of the present distribution, i.e. [M/H]$= -0.07$,
which is very similar to the result of
\citet{luck05} who found $\langle$[Fe/H]$\rangle = -0.04$ for a sample
of 114 thin-disk stars within 15 parsecs of the sun.

\section{Concluding Remarks}

We have presented results for 1676 dwarf and giant stars within 40pc
of the Sun including new, homogeneous spectral types, basic physical
parameters, and measures of chromospheric activity.  We will complete
our study of the dwarf and giant stars earlier than M0 within 40pc in
the third and final paper of this series.  The goals of this project
are to characterize the stellar population in the solar neighborhood
and to provide data that will be useful in the selection of targets
for the {\it Space Interferometry Mission} and the {\it Terrestrial
  Planet Finder} mission.  

The data presented in this paper are currently available on the
project's Web site, and work is continuing on the remaining stars in
the project.  

This work has been carried out under contract with NASA/JPL (JPL
contract 526270) and has been partially supported by grants from the
Vatican Observatory and Appalachian State University.  This research
made use of the SIMBAD database, operated at CDS, Strasbourg, France.
We would also like to thank Jean-Claude Mermilliod for his assistance
in the compilation of photometry for our database, and his maintenance
(with M. Mermilliod)
of the web-based General Catalogue of Photometric Data
\citep{mermilliod97} which has proved very useful in this research.
Many thanks to Kelly Kluttz and Chris Jackolski, both of Appalachian
State University for help in various aspects of this project.

\include{tab2}

\include{tab5}

\end{document}

%% file: tab5.tex
\begin{deluxetable}{rrr}
\tablewidth{0pt}
\tabletypesize{\scriptsize}
\tablenum{5}
\tablewidth{0pt}
\tablecaption{Solar Analogues}
\tablehead{\colhead{HIP} & \colhead{HD} & \colhead{Notes}} 
\startdata
   699 & 361 & * \\
  1444 & 1388 & D  \\
  1954 & 2071 & *D \\
  3578 & 4392 &   \\
  6455 & 8406 & * \\
  7822 & 10370 &   \\
 14501 & 19467 & *D \\
 18844 & 25874 & **D \\
 26394 & 39091 & EP \\
 27058 & 38277 & * \\
 27244 & 38973 & D  \\
 30104 & 44594 & **D \\
 30476 & 45289 & **D \\
 30503 & 45184 & **D  \\
 34879 & 55693 & *D  \\
 36512 & 59967 & ** \\
 39330 & 66653 & ** \\
 40283 & 68978 & D  \\
 41317 & 71334 & **D \\
 42291 & 73524 & D  \\
 44713 & 78429 & **D \\
 50075 & 88742 & D  \\
 53837 & 95521 & *  \\
 54287 & 96423 & D  \\
 55900 & 99610 & D  \\
 60370 & 107692 & *D  \\
 60729 & 108309 & *D  \\
 62345 & 111031 & D  \\
 66047 & 117618 & D  \\
 70459 & 125881 & D  \\
 74273 & 134060 & D  \\
 74389 & 134664 & ** \\
 77740 & 141937 & EP \\
 78169 & 142415 & D  \\
 79578 & 145825 & **D \\
 79658 & 146070 & * \\
 86991 & 160859 & *  \\
 89042 & 165499 &   \\
 90223 & 168871 & D  \\
 91287 & 171665 & D  \\
 94154 & 177409 &   \\
 96901 & 186427 & *D  \\
 97405 & 186853 &  \\
 98589 & 189625 & **D  \\
 98621 & 188748 &  \\
 98813 & 189931 & *  \\
 98959 & 189567 & **D \\
101905 & 196390 & D  \\
102128 & 196068 & D \\
103654 & 199190 & D \\
105184 & 202628 & **D \\
105214 & 202457 &  \\
106213 & 204385 & D  \\
107620 & 207043 &   \\
107649 & 207129 & D  \\
108158 & 207700 & D  \\
109821 & 210918 & **D  \\
110712 & 212168A &  D  \\
113357 & 217014 & EP  \\
114424 & 218730 &  D  \\
117066 & 222669 & *  \\
\enddata
\tablecomments{Symbols in the notes column have the following
  meanings: *; A solar analogue with spectral type within 1 subclass
  of the sun.  **; A candidate solar twin with spectral type, and
  physical parameters very similar to that of the sun.  D;  A star
  which is already in the Keck, Lick \& AAT doppler planet-search
  program (see Valenti \& Fischer 2005). EP; A star with
  a known exoplanet.}

\end{deluxetable}